\documentclass[conference]{IEEEtran}

\usepackage{amsthm}
\usepackage[cmex10]{amsmath}
\usepackage{graphicx}
\usepackage[tight,footnotesize]{subfigure}
\usepackage{amsfonts}
\usepackage{algorithmic}
\usepackage{algorithm}
\usepackage{graphicx}
\usepackage{subfigure}
\usepackage{bm}
\usepackage{times,color,amssymb,epsfig,psfrag,stfloats,enumerate}

\ifCLASSINFOpdf
\else
\fi
\begin{document}
%
\title{Proactive Push with Energy Harvesting Based Small Cells in Heterogeneous Networks}

\author{\IEEEauthorblockN{Jie Gong\IEEEauthorrefmark{1}, Sheng Zhou\IEEEauthorrefmark{1}, Zhenyu Zhou\IEEEauthorrefmark{2}, Zhisheng Niu\IEEEauthorrefmark{1}\\}
\IEEEauthorblockA{\IEEEauthorrefmark{1}Tsinghua National Laboratory for Information
Science and Technology,\\
Department of Electronic Engineering, Tsinghua University, Beijing 100084, China}
\IEEEauthorblockA{\IEEEauthorrefmark{2}State Key Laboratory of Alternate Electrical Power System with Renewable Energy Sources,\\
School of Electrical and Electronic Engineering, North China Electric Power University, Beijing 102206, China\\
Email: gongj13@mail.tsinghua.edu.cn}
}

\maketitle

\begin{abstract}
Motivated by the recent development of energy harvesting communications, and the trend of multimedia contents caching and push at the access edge and user terminals, this paper considers how to design an effective push mechanism of energy harvesting powered small-cell base stations (SBSs) in heterogeneous networks. The problem is formulated as a Markov decision process by optimizing the push policy based on the battery energy, user request and content popularity state to maximize the service capability of SBSs. We extensively analyze the problem and propose an effective policy iteration algorithm to find the optimal policy. According to the numerical results, we find that the optimal policy reveals a state dependent threshold based structure. Besides, more than 50\% performance gain is achieved by the optimal push policy compared with the non-push policy.
\end{abstract}

\IEEEpeerreviewmaketitle

\section{Introduction}
Due to the rapidly growing multimedia traffic over the air and the critical concern regarding
$\mathrm{CO}_2$ emissions, green wireless communications are urgently required. There have been some candidate technologies which are demonstrated as effective ways to achieve green wireless access, such as energy harvesting (EH), multicast and heterogeneous networks. EH technology \cite{ozel2011transmission, gunduz2014designing}, which utilizes the energy from natural sources such as solar, wind, and kinetic activities, can greatly reduce the wireless communication power consumption from the conventional power supply, i.e., power grid. Wireless multicast \cite{niu2008new, liu2013utility} holds the promise of achieving huge energy efficiency gain via delivering commonly interested multimedia contents to multiple users simultaneously by broadcasting a single data stream to different users, which avoids duplicated retransmissions of the same content. Heterogeneous networks provide higher data rate to users by cutting down the distance between users and base stations (BSs) with densely deployed small-cell BSs (SBSs). However, each technology has its limitations based on the state of the art. Because of the limited battery capacity, energy waste or shortage will occur when energy and traffic arrivals mismatch with each other. On the other hand, to enable wireless multicast, some user requests need be delayed to wait for concurrent transmission, which may severely damage the quality of service (QoS) of the earlier demands. Finally, the deployment of SBSs is not flexible as it may cause high cost for deploying the supporting power lines and high-speed backhaul links.

To break through the limitations for higher energy efficiency, we introduce the proactive push mechanism \cite{podnar2002distributed} to combine the technologies mentioned above. Powering the SBSs with EH devices greatly increases the flexibility of heterogeneous network deployment. Based on the EH status and content popularity distribution, the SBSs proactively cache and push the contents earlier than the actual demands. In reward, the time duration in which the desired content can be delivered is greatly extended, so that the delivery can flexibly match to the EH process. On the other hand, from the energy point of view, as the harvested energy can be effectively and timely used, the energy waste due to the battery capacity limitation can be avoided. In other words, proactive push is a novel way of information and power transfer over the hyper dimension of space (small cell to users) and time (now to the future) respectively, which is different from the joint transfer over space only \cite{zhou2013wireless}.

Proactive push is supported by the recent trends on the development of last-mile wireless access hardwares and mobile devices. To reduce the core network overhead and enhance user experiences in terms of delay and rate, contents are suggested to be cached at the SBSs \cite{golrezaei2013femtocaching, Shanmugam2013femtocaching} or relay nodes \cite{wang2011on}, with proactive caching schemes \cite{bastug2014living}. Also, there have been some commercial products such as HiWiFi \cite{Hiwifi} with large storage for caching. On the other hand, with the rapid improvement of data storage capacity, user devices are capable of storing large amount of data for potential user requests. And the network capacity gain provided by proactive push in the integrated broadcast and communication network is analyzed in Ref.~\cite{wang2014push}. With the large user storage capacity and the available contents at the edge nodes, proactive push by EH powered SBSs is considered to be practical and effective.

Recently, EH based SBSs are used to cache contents \cite{sharma2013greencache} for the deployment flexibility and energy consumption reduction, and the GreenDelivery framework for content delivery with EH powered small cells is proposed in \cite{Zhou2014greendelivery}. As far as we know, the proactive push optimization is still an open problem in EH powered SBSs. And the problem is not trivial since it needs to jointly consider energy state, traffic load as well as content popularity. Pushing a content to a set of users typically consumes more energy than unicasting a required content to a single user as push needs to guarantee the data rate of the worst-channel user. While the more contents are pushed, the fewer unicast requests are generated since more contents can be found in users' local storage. Hence, there is a tradeoff between high energy consumption and low request generation rate by push mechanism which needs extensive study.

In this paper, we try to optimize the proactive push policy of a EH powered SBS in heterogeneous wireless networks. The objective is to minimize the ratio of user requests handled by the macro BS which happens when the SBS is of low energy or is pushing another content. We formulate the problem using Markov decision process (MDP) \cite{bertsekas2005dynamic} tool with detailed modeling of state, action, cost function and state transition probability, and find the optimal stationary policy via policy iteration algorithm. Numerical results are provided to illustrate the structure of optimal policy and the performance gain compared with non-push policy.

The rest of the paper is organized as follows. Section \ref{model} presents the system model. The problem is formulated and analyzed in Section \ref{sec:problem}. Some numerical results are provided in Section \ref{simulation} for performance evaluation. Finally, Section \ref{concl} concludes the paper.

\section{System Model} \label{model}
\begin{figure}
\centering
\includegraphics[width=3.5in]{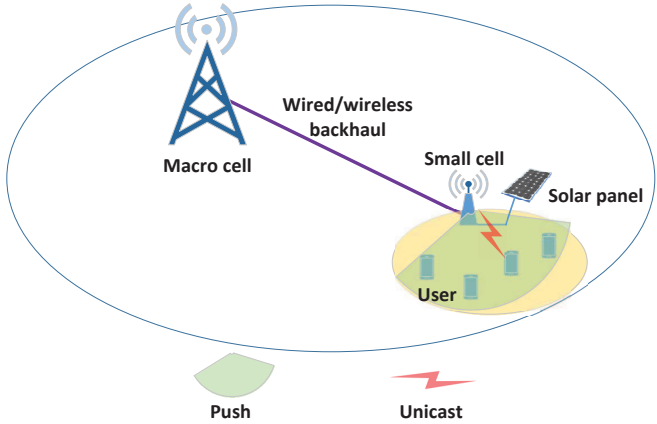}
\caption{Two-tier heterogeneous cellular network. There are multiple small cells in a macro cell. Only one of them is depicted as we focus on a single small-cell analysis.} \label{fig:system}
\end{figure}

We consider a second-tier small-cell with radius $R$ in a two-tier heterogeneous cellular network as shown in Fig.~\ref{fig:system}. The SBS is powered by renewable energy solely, and the harvested energy can be stored in a battery with finite capacity $B_{\mathrm{max}}$. There is only one frequency channel for data transmission in each small cell, and the SBS has a high-speed wired/wireless backhaul link to the macro-cell BS to fetch any content immediately when required. When the SBS has sufficient energy, it can either unicast a required content to the specific user who requires it, or multicast a popular content to all the users in its coverage, i.e., push. When the battery energy is not enough, the BS enters into sleep mode and the content request will be handled by the macro BS. The SBS can also choose to sleep even though the battery is sufficient for transmission. In this way, there will be more energy available in the later times. At the user side, if a required content is in the users cache, it can directly access the content and does not need to trigger a transmission from the SBS or the macro BS. In this paper, we assume each user has sufficient caching capacity so that any pushed contents can be successfully stored, and focus on how to design push policy to fully utilize the renewable energy in SBSs.

Assume there are a total of $N$ contents of equal length that the users are interested in. Each content has a minimum average data rate requirement $r_0$. Hence, the content transmission time is identical for all contents if they are transmitted with rate $r_0$. Then the system is slotted with the length of each period equal to the content transmission time, denoted by $T_p$. The popularity of the contents varies from each other. Statistical researches have shown that the content popularity distribution is well fitted by the Zipf distribution \cite{cha2007tube, golrezaei2013femtocaching}. Specifically, the popularity of the $i$-th ranked content among the $N$ contents can be expressed as
\begin{equation}
f_i = \frac{1/i^v}{\sum_{j=1}^N 1/j^v}, \label{eq:pop}
\end{equation}
where $v \ge 0$ is the skew parameter. In the real network, people are more interested in the contents with higher popularity, which will result in higher request probability. In addition, as people's interest changes over time, some contents may be outdated and replaced by new ones. We assume in each period the probability that a piece of content leaves the system and is replaced by a new one is $p_c \in [0, 1]$. The leaving content is randomly chosen among all the contents $1, 2, \ldots, N$.

In each period, there is a content request with probability $p_u\in [0, 1]$. And the user generating the request is assumed uniformly distributed in the small cell. The channel model considers large-scale pathloss effect as well as small-scale fast fading. For each content transmission, the data rate can be calculated as
\begin{equation}
r = \mathbf{E}_h\left[W\log_2\left(1+\frac{P_t|h|^2\beta d^{-\alpha}}{\sigma^2+I}\right)\right], \label{eq:rate}
\end{equation}
where $W$ is the bandwidth of the SBS, $P_t$ is the transmit power, $h$ is the small-scale fast fading coefficient, $\beta$ and $\alpha$ represent the pathloss constant and the pathloss exponent, respectively, $d$ is the transmission distance, $\sigma^2+I$ is the noise plus interference power. Assume the SBSs and the macro BS are allocated with orthogonal frequency bands. As a result, there is no inter-tier interference, and the interference is only caused by the randomly and densely deployed SBSs working in the same frequency band. Hence, according to the law of large numbers, the noise plus interference together can be considered as additive white Gaussian noise (AWGN) with variance $\sigma^2+I$. $\mathbf{E}_h$ is the expectation operator with respect to $h$. Based on the channel model, the required power for sending a content to a user with distance $d$ can be obtained by setting $r=r_0$ and solving (\ref{eq:rate}) numerically.

\begin{figure}[th]
\centering
\includegraphics[width=3.5in]{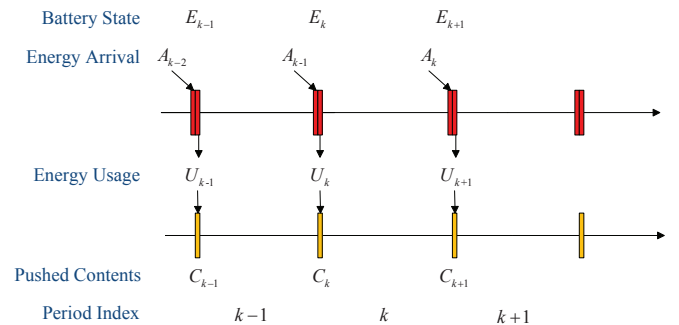}
\caption{Timeline of the slotted system.} \label{fig:queue}
\end{figure}

Next, we describe the slotted system model in detail. As illustrated in Fig.~\ref{fig:queue}, at the beginning of the period indexed by $k$, the battery energy is denoted by $E_k$, and the number of pushed contents is denoted by $C_k$. In our analysis, we always push the most popular contents to the users. According to this simple push policy, the pushed contents are those ranked from 1 to $C_k$. Based on the situation that whether there is a user request or not, whether the requested content is pushed or not, and how much energy is required for unicasting the required content, the BS decides its action, i.e., unicast the required content, push a content, or sleep. Then at the beginning of the next period $k+1$, the battery energy state is updated as
\begin{equation}
E_{k+1} = \min\{B_{\mathrm{max}}, E_k-U_k+A_k\}, \label{eq:Ek}
\end{equation}
where $U_k$ is the energy used for transmission which satisfies $U_k \le E_k$, and $A_k$ is the amount of harvested energy in period $k$, which is assumed i.i.d. If the BS decides to sleep, $U_k=0$. $C_{k+1}$ takes values as
\begin{equation}
C_{k+1} \in \{\max\{0, C_k-1\}, C_k, \min\{N, C_k+1\}\} \label{eq:Ck}
\end{equation}
according to the BS's action and the content update behavior. At this moment, the BS takes its action based on the renewed system status. When a user requests a content that is not in its cache, but the BS decides not to unicast for some reason, it needs to be handled by the macro BS, which causes additional energy and resource allocation from the macro BS. Intuitively, the ratio of user requests handled by the macro BS indicates the harvested energy utilization efficiency. The lower ratio of user requests handled by the macro BS is achieved, the more efficiently the harvested energy is used. In the next section, we will provide the problem formulation aiming at minimizing the ratio.

\section{Problem Formulation and MDP Solution} \label{sec:problem}
Our problem can be described as minimizing the ratio of user requests handled by the macro BS over the total user requests by adjusting the behavior of the SBS under the energy constraint. Mathematically, the objective can be expressed as
\begin{equation}
\min \lim_{K\rightarrow +\infty} \frac{\bar{K}}{K}, \label{eq:obj}
\end{equation}
where $\bar{K}$ is the number of user requests handled by the macro BS and $K$ is the number of total periods. Notice that the objective function in (\ref{eq:obj}) is not the ratio of user requests handled by the macro BS over the total user requests but related with it as
\begin{equation}
\frac{\bar{K}}{K} = \frac{\bar{K}}{\tilde{K}}\frac{\tilde{K}}{K} = \frac{\bar{K}}{\tilde{K}}p_u,
\end{equation}
where $\tilde{K}$ is the total number of user requests during $K$ time periods, and recall that $p_u$ is the content request probability. For a given $p_u$, minimizing the ratio of user requests handled by the macro BS over the total user requests is equivalent with (\ref{eq:obj}).

To solve the problem (\ref{eq:obj}), we need to decide the BS's action in each period. As the per-period action is determined based on the system state at the beginning of each period, the problem can be modeled as a MDP optimization problem \cite{bertsekas2005dynamic}. MDP, also termed as dynamic programming (DP) \cite{bertsekas2005dynamic}, is an effective tool and widely used for the control optimization of stochastic process. It deals with the set of problems with controlled Markov process where the control action in each stage\footnote{In this paper, the term ``stage" is equivalent with the term ``period".} is based only on the current system state. A standard MDP problem contains the following elements: state, action, cost function, and state transition. Next, we re-formulate our problem as a MDP optimization problem by describing the elements one by one.

\subsubsection{System State}
The state of the system in stage $k$ is denoted by
\begin{equation}
x_k = (E_k, Q_k, C_k),
\end{equation}
where as mentioned before, $E_k$ and $C_k$ is the battery energy and the number of pushed contents, respectively. $Q_k$ is the user request state. We set $Q_k=0$ if there is no user request or the requested content is already in the user's cache. Otherwise, $Q_k$ represents the energy consumption for completing the required content transmission. For a user request generated at distance $d$, $Q_k = P_t(d)T_p$, where $P_t(d)$ is the transmission power obtained by solving (\ref{eq:rate}) with $r = r_0$. Denote the state space as $\mathcal{S}$.

As energy and user locations take continuous values, there is a continuous state space, which makes the problem difficult to be solved. So we further discretize the state space $\mathbf{S}$ into a finite set to make the problem tractable. The energy is discretized with unit energy $E_{\mathrm{unit}}$. Then the energy state is $E_k \in \{0, 1, \ldots, E_{\mathrm{max}}\}$ with $E_{\mathrm{max}}E_{\mathrm{unit}} = B_{\mathrm{max}}$. $E_k = i$ corresponds to $iE_{\mathrm{unit}}$ amount of energy, and similarly for energy arrival $A_k$. To discretize $Q_k$, we select a series of distances $0<d_1 < d_2 < \ldots < d_M = R$ so that $P_t(d_i)T_p = l_iE_{\mathrm{unit}}$ where $l_i$ is a positive integer for any $i = 1, 2, \ldots, M$. For any user with distance to the BS ranging from $d_{i-1}$ to $d_{i}$, we unicast the required content with energy $P_t(d_i)T_p$, which guarantees the minimum data rate $r_0$ for all the users in this area. And we set $l_0=0$ denoting that the required energy for unicast is zero. Then we have $Q_k \in \{0, 1, \ldots, M\}$, where $Q_k = i$ corresponds to the case that $l_iE_{\mathrm{unit}}$ amount of energy is required for unicasting the content.

With the discretization procedure, the state space $\mathbf{S}$ is of dimension $(E_{\mathrm{max}}+1)\times (M+1) \times (N+1)$.

\subsubsection{BS Action}
The SBS has three actions to choose: sleep, unicast the required content, and push the most popular un-pushed content. We define the action takes values in set $\mathcal{U} = \{0, 1, 2\}$ as
\begin{equation}
u_k = \left\{ \begin{array}{ll} 0, & \textrm{sleep} \\
1, & \textrm{unicast~the~required~content}\\
2, & \textrm{push~the~most~popular~un-pushed~content}\end{array}\right.
\end{equation}

Notice that in different states, the BS may not be able to take all the three actions. A simple example is that if $E_k = 0$, the BS can do nothing but sleep, i.e., $u_k = 0$. Hence, the action space is state-dependent, which can be expressed as $u_k \in \mathcal{U}_k(x_k)$. If $E_k \ge Q_k$, $ 1 \in \mathcal{U}_k(x_k)$, i.e., the energy for unicast can be satisfied. To push a content, it must be guaranteed that all the users in the small cell coverage can receive the content with rate $r_0$. So the user at cell edge (distance to the BS is $R$) must be covered. Then we conclude that if $E_k \ge l_M$, $ 2 \in \mathcal{U}_k(x_k)$, i.e., the energy for push can be satisfied.

\subsubsection{Cost Function}
The cost function depends on both the system state and the action, hence is denoted by $g_k(x_k, u_k)$. In our problem, the cost happens if and only if there is a user request handled by the macro BS. Hence, we have $g_k(x_k, u_k) \in \{ 0, 1 \}$. $g_k(x_k, u_k) = 1$ if the user request is handled by the macro BS, and $g_k(x_k, u_k) = 0$ otherwise. Mathematically, we can express it as
\begin{equation}
g_k(x_k, u_k) = \left\{ \begin{array}{ll} 1, & \textrm{if~} Q_k > 0, u_k \neq 1 \\
0, & \textrm{otherwise}\end{array}\right.
\end{equation}

\subsubsection{State Transition}
The state transition is expressed as the conditional probability
\begin{align}
&p_{x_k\rightarrow x_{k+1}|u_k} \nonumber\\
= &\mathrm{Pr}(E_{k+1}, Q_{k+1}, C_{k+1}|E_k, Q_k, C_k, u_k)\nonumber\\
= &\mathrm{Pr}(E_{k+1}|E_k, Q_k, u_k)\mathrm{Pr}(C_{k+1}|C_k, u_k)\mathrm{Pr}(Q_{k+1}|C_{k+1}), \label{eq:statetrans}
\end{align}
where the second equality is derived based on the law of total probability and the fact that for the given action $u_k$, $E_{k+1}$ only depends on $E_k$ and $Q_k$ according to (\ref{eq:Ek}), and $C_{k+1}$ only depends on $C_k$ according to (\ref{eq:Ck}). While $Q_{k+1}$ depends on $C_{k+1}$ because $C_{k+1}$ decides the probability with which a content has been pushed, hence influences the probability with which a unicast is required.

We calculate the state transition probability according to (\ref{eq:statetrans}). Firstly, to calculate the energy state transition probability, we denote $p_a(i), i = 0, 1, \ldots$ as the probability that $iE_{\mathrm{unit}}$ amount of energy is arrived, which satisfies $p_a(i) \in [0, 1], \sum_i p_a(i) = 1$. To simplify the description, we set $p_a(i) = 0, \forall i = -1, -2, \ldots$. Then we have
\begin{align}
&\mathrm{Pr}(E_{k+1}|E_k, Q_k, u_k) = \nonumber\\
&\left\{ \begin{array}{l}
p_a(E_{k+1} - E_k),  \quad\qquad \textrm{if~} u_k = 0, E_{k+1} < E_{\mathrm{max}}\\
1-\sum\limits_{i=0}^{E_{\mathrm{max}}-E_k-1}p_a(i), \quad \textrm{if~} u_k = 0, E_{k+1} = E_{\mathrm{max}}\\
p_a(E_{k+1}-E_k+l_{Q_k}), \\
\qquad\quad\quad \textrm{if~} 0 < l_{Q_k} \le E_k, u_k  = 1, E_{k+1} < E_{\mathrm{max}}\\
1-\sum\limits_{i=0}^{E_{\mathrm{max}}-E_k+l_{Q_k}-1}p_a(i), \\
\qquad\quad\quad \textrm{if~} 0 < l_{Q_k} \le E_k, u_k  = 1, E_{k+1} = E_{\mathrm{max}}\\
p_a(E_{k+1}-E_k+l_{M}), \\
\qquad\qquad\qquad \textrm{if~} l_{M} \le E_k, u_k  = 2, E_{k+1} < E_{\mathrm{max}}\\
1-\sum\limits_{i=0}^{E_{\mathrm{max}}-E_k+l_{M}-1}p_a(i), \\
\qquad\qquad\qquad \textrm{if~} l_{M} \le E_k, u_k  = 2, E_{k+1} = E_{\mathrm{max}}\\
\end{array}\right.
\end{align}

Note that the action $u_k = 0$ can be taken in any states, while $u_k = 1$ can be taken under the condition that $0 < l_{Q_k} \le E_k$, and $u_k = 2$ with condition $l_{M} \le E_k$. Also note that when $E_{k+1} = E_{\mathrm{max}}$, the energy arrival may exceed the battery capacity. So the probability is calculated by summarizing all the possible energy arrival conditions.

Secondly, as in each stage, at most one content is pushed to users, and also at most one content will be replaced by a new one, $C_k$ can only transit to its neighboring values $C_k+1, C_k-1$ or keeps constant. The pushed content state is updated as
\begin{align}
&\mathrm{Pr}(C_{k+1}|C_k, u_k) = \nonumber\\
&\left\{ \begin{array}{ll}
p_c\frac{C_k}{N}, & \textrm{if~} u_k < 2, C_{k+1} = C_{k}-1\ge 0\\
1-p_c\frac{C_k}{N}, & \textrm{if~} u_k < 2, C_{k+1} = C_{k}\\
1-p_c\frac{C_k}{N}, & \textrm{if~} u_k = 2, C_{k+1} = C_{k}+1\le N\\
p_c\frac{C_k}{N}, & \textrm{if~} u_k = 2, C_{k+1} = C_{k}<N\\
0. & \textrm{else}
\end{array}\right.
\end{align}

Note that when a pushed content is replaced by a new one, it is removed from users' cache. While when the replaced content is not pushed at all, there is no influence to $C_k$.

Finally, the user request state transition is
\begin{align}
&\mathrm{Pr}(Q_{k+1}|C_{k+1}) = \nonumber\\
&\left\{ \begin{array}{ll}
(1-p_u) + p_uf_{C_{k+1}}, & \textrm{if~} Q_{k+1} = 0\\
p_u(1-f_{C_{k+1}})\frac{d_{m}^2-d_{m-1}^2}{R^2}, & \textrm{if~} Q_{k+1} = m > 0
\end{array}\right.
\end{align}
where $f_{C_{k+1}}$ is calculated according to (\ref{eq:pop}) and $d_0=0$. $Q_{k+1} = 0$ means that either there is no user request generated or the user request can be satisfied by caching, i.e., the required content has been pushed. Otherwise, as the users are assumed uniformly distributed in the cell, the request is generated with distance to BS ranging from $d_{m-1}$ to $d_m$ with the probability equal to the ratio of the circular ring area to the cell area.

\subsection{MDP Problem Formulation and Optimization}
Based on the above MDP-based system modeling, the original optimization problem (\ref{eq:obj}) can be re-written as
\begin{equation}
\min \lim_{K \rightarrow +\infty}\frac{1}{K}\mathbf{E}\left[ \sum_{k=0}^{K-1}g(x_k, u_k(x_k))\right]. \label{eq:MDPobj}
\end{equation}
The expectation operation is taken over all the random parameters including energy arrival, user request, and content update. The optimization is taken over all the possible policies $\{u_1, u_2, \ldots\}$. It can be proved that for any two states, there is a stationary policy $\bm{u}$ so that one state can be accessed with non-zero probability from the other with finite steps. Consequently, the optimization is irrelevant with the initial state $x_0$, and there exists an optimal stationary policy $\bm{u}^*$ \cite[Sec~4.2]{bertsekas2005dynamic}.

According to \cite[Prop.~4.2.1]{bertsekas2005dynamic}, the optimal average cost $\lambda^*$ together with some vector $\bm{h}^* = \{h^*(x)|x \in \mathcal{S}\}$ satisfies the Bellman's equation
\begin{equation}
\lambda^* + h^*(x) = \min_{u\in \mathcal{U}(x)}\left[ g(x, u) + \sum_{y \in \mathcal{S}}p_{x\rightarrow y | u}h^*(y)\right]. \label{eq:bellman}
\end{equation}
Further more, if $u^*(x)$ attains the minimum value of (\ref{eq:bellman}) for each $x$, the stationary policy $\bm{u}^*$ is optimal. Based on the Bellman's equation, instead of the long term average cost minimization, we only need to deal with (\ref{eq:bellman}) which only relates with per-stage cost $g(x, u)$ and state transition $p_{x\rightarrow y | u}$. The policy iteration algorithm \cite[Sec.~4.4]{bertsekas2005dynamic} can effectively solve the problem, which will be detailed in the next subsection.

\begin{figure*}
  \centering
  \subfigure[$C_k = 0$]{
    \label{subfig:C0}
    \includegraphics[width=2 in]{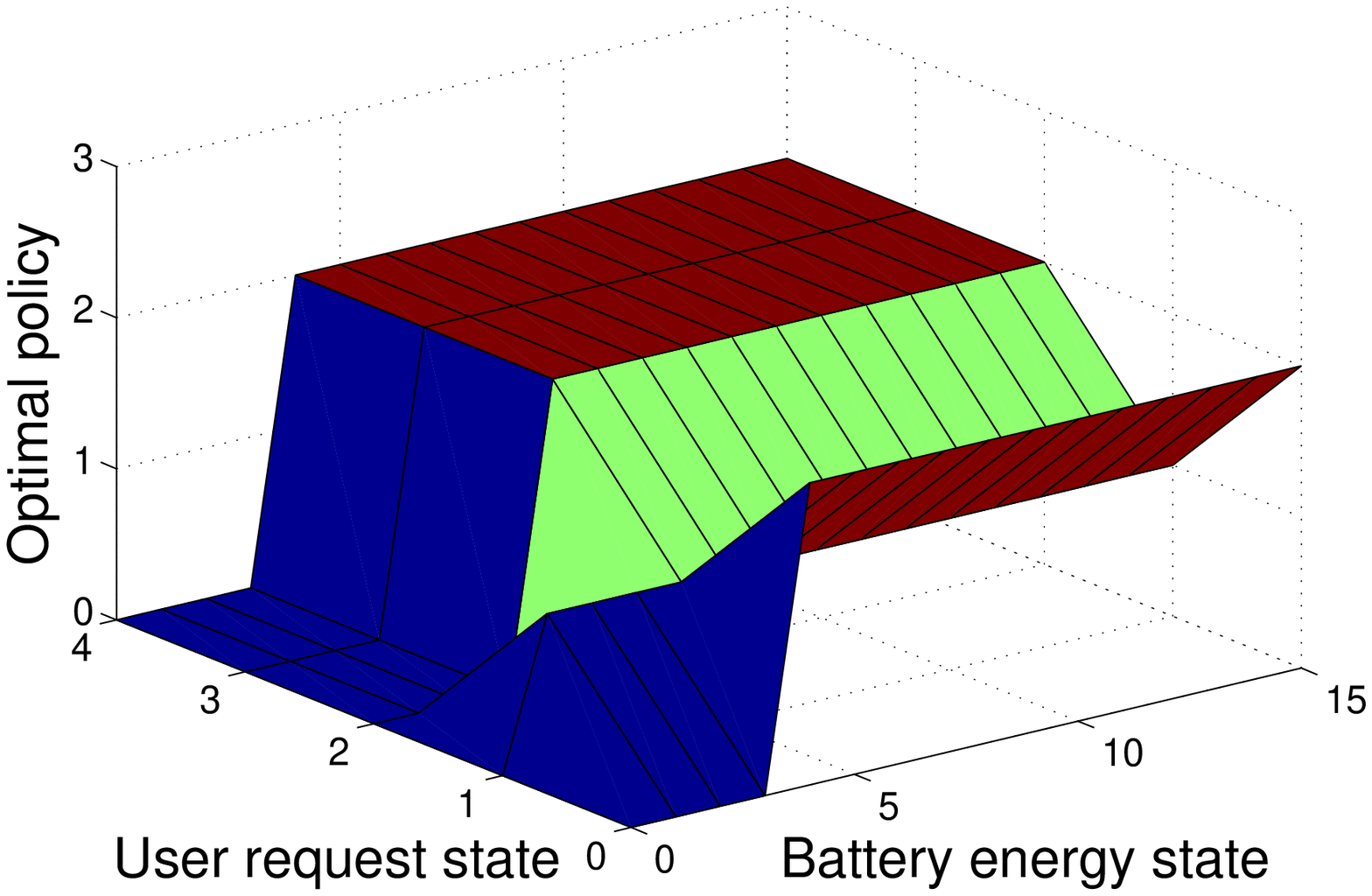} }
  \subfigure[$C_k = 4$]{
    \label{subfig:C4}
    \includegraphics[width=2 in]{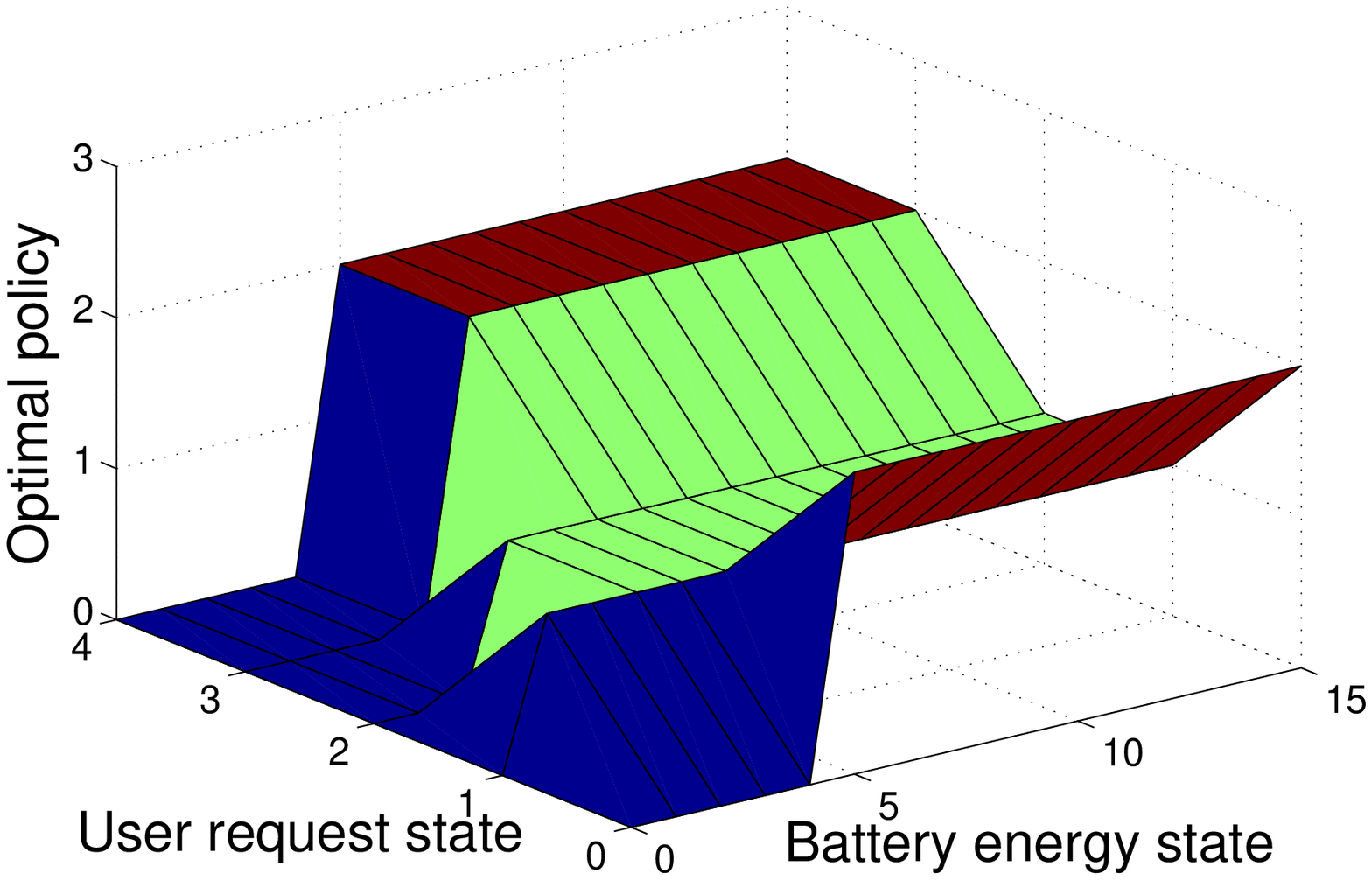} }
  \subfigure[$C_k = 9$]{
    \label{subfig:C9}
    \includegraphics[width=2 in]{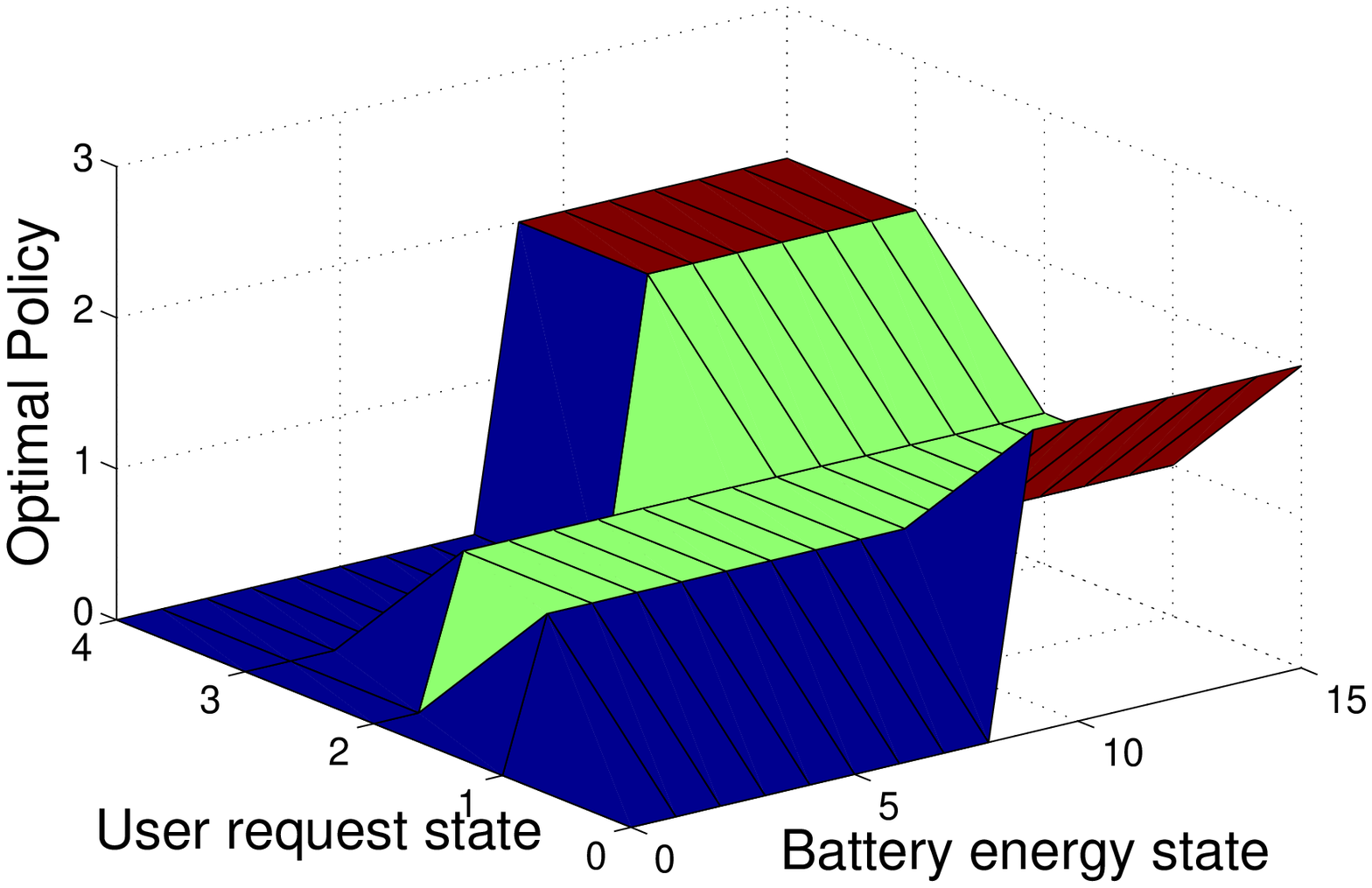} }
  \subfigure[$C_k = 13$]{
    \label{subfig:C13}
    \includegraphics[width=2 in]{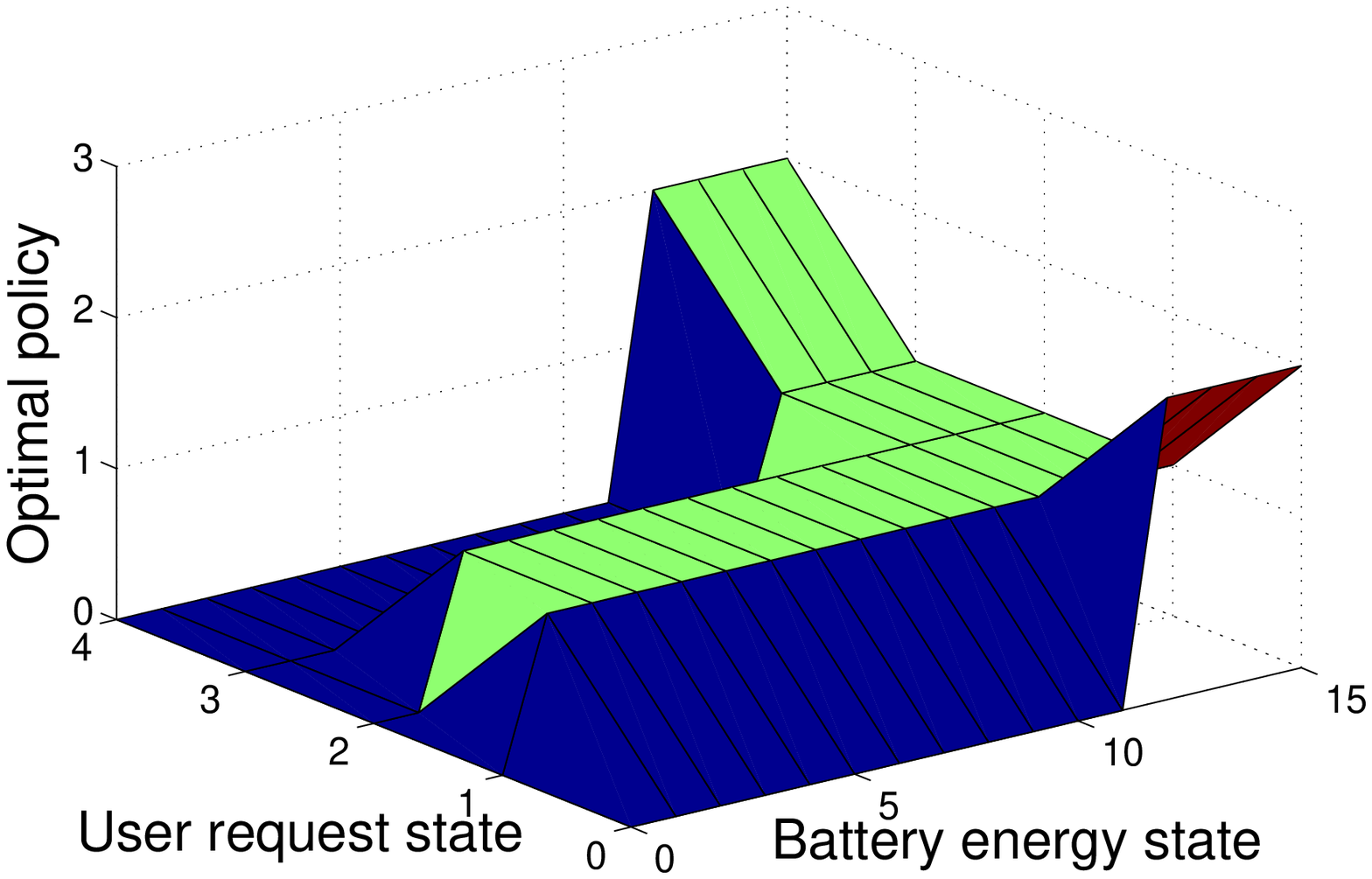} }
  \subfigure[$C_k = 19$]{
    \label{subfig:C19}
    \includegraphics[width=2 in]{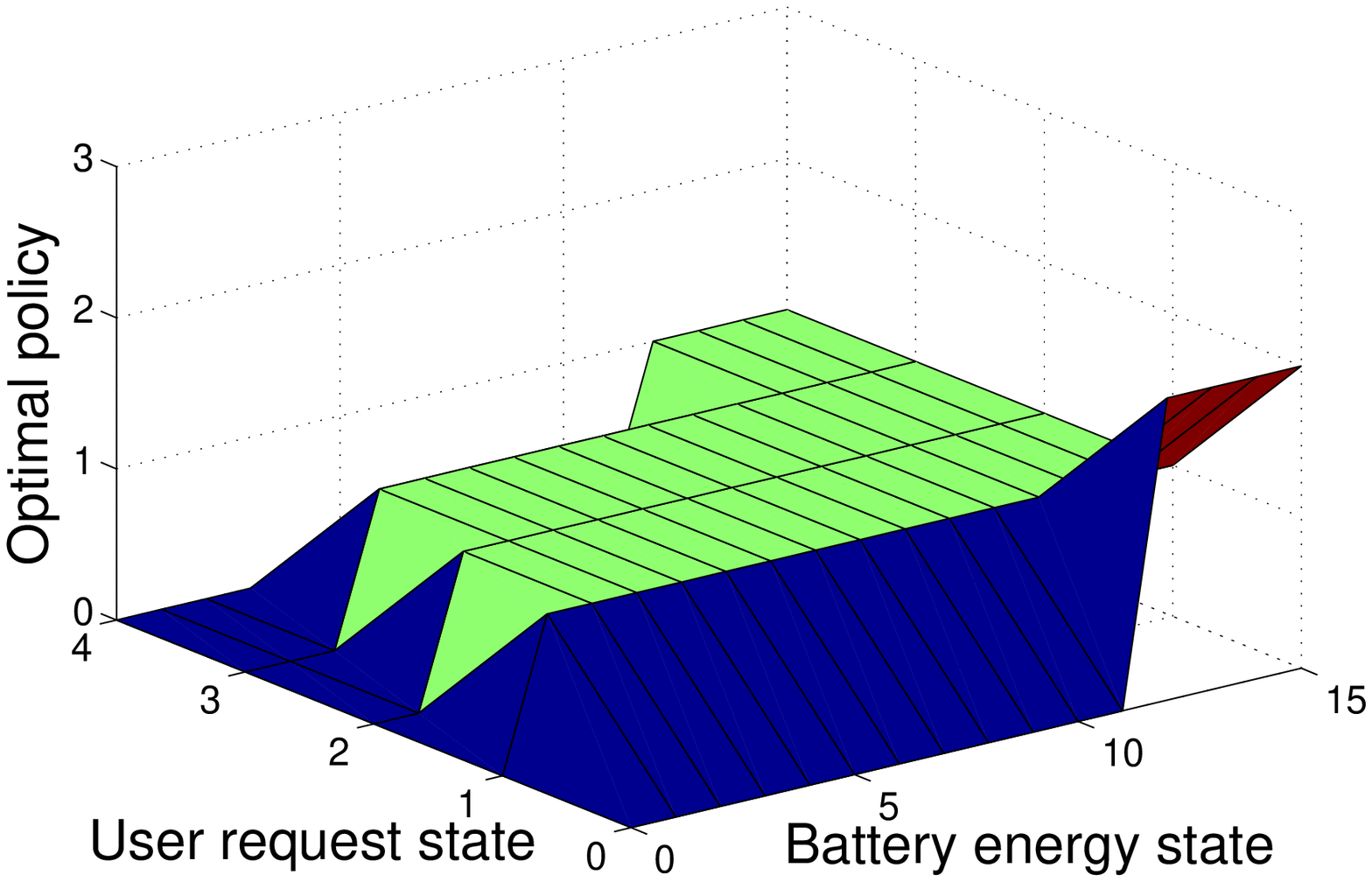}}
  \subfigure[$C_k = 20$]{
    \label{subfig:C20}
    \includegraphics[width=2 in]{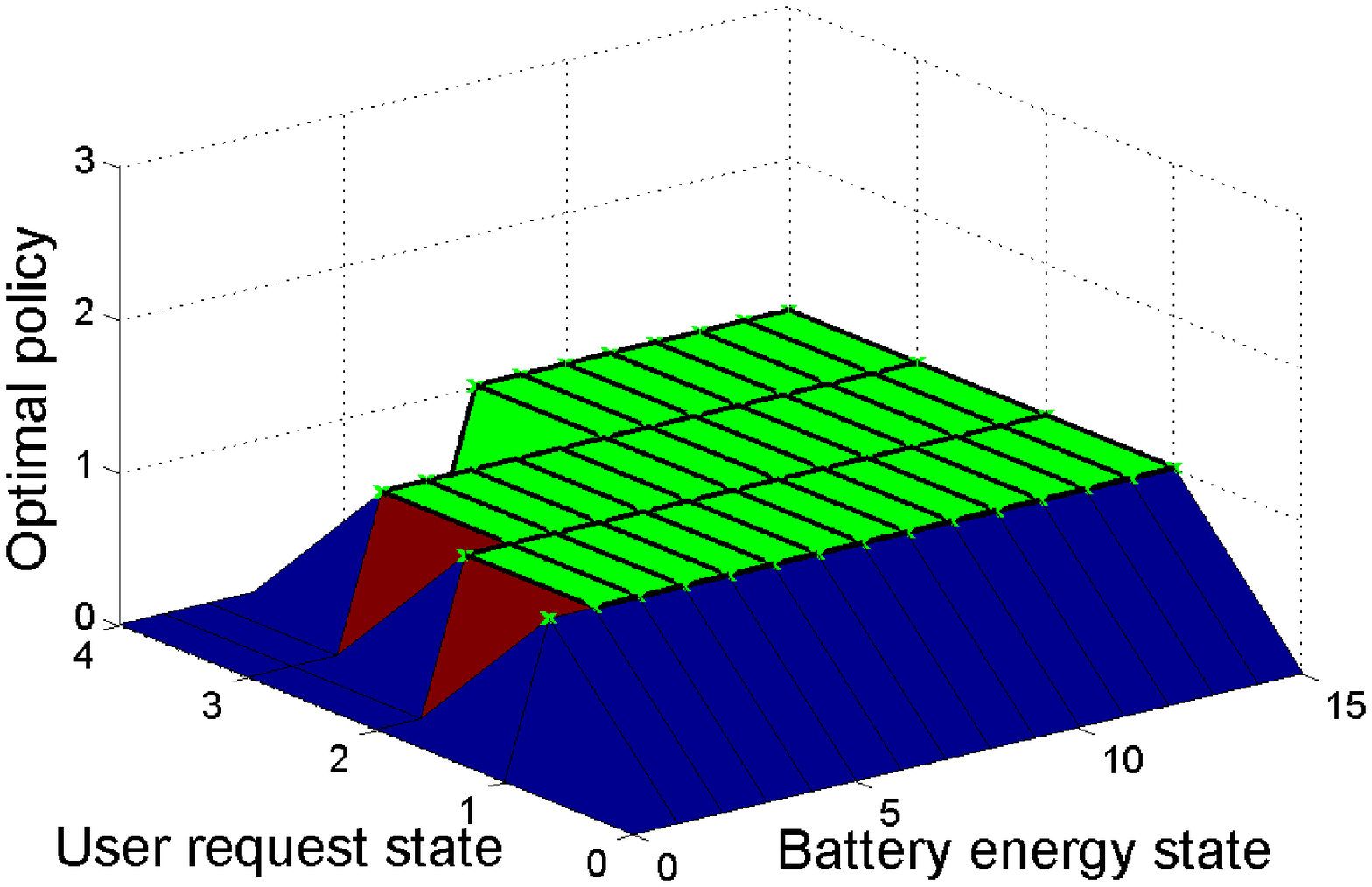} }
\caption{Optimal policy sampled w.r.t.~pushed content state with parameter $p_c = 0.3, p_u = 0.7, \bar{a} = 0.8$.}
\label{fig:Optpolicy}
\end{figure*}

\subsection{Policy Iteration Algorithm}

The policy iteration algorithm starts with any feasible stationary policy, and improves the objective step by step. Suppose in the $j$-th step, we have a stationary policy denoted by $\bm{u}^{(j)}$. Based on this policy, we perform \emph{policy evaluation} \cite[Sec.~4.4]{bertsekas2005dynamic} step, i.e., we solve the following linear equations
\begin{equation}
\lambda^{(j)} + h^{(j)}(x) = g(x, u^{(j)}(x)) + \sum_{y \in \mathcal{S}}p_{x\rightarrow y | u^{(j)}(x)}h^{(j)}(y) \label{eq:linear}
\end{equation}
for $\forall x \in \mathcal{S}$ to get the average cost $\lambda^{(j)}$ and vector $\bm{h}^{(j)}$. Notice that there are $(E_{\mathrm{max}}+1)\times (M+1) \times (N+1)$ equations but $(E_{\mathrm{max}}+1)\times (M+1) \times (N+1)+1$ unknown parameters, hence more than one solutions exist, which are different with each other by a constant value for all $h^{(j)}(x)$. Without loss of generality, we can set for example \begin{equation}
h^{(j)}(E_{\mathrm{max}}+1, M+1, N+1)=0, \label{eq:hconstr}
\end{equation}
then the solution for (\ref{eq:linear}) is unique.

As $\bm{u}^{(j)}$ may not be the optimal policy, we subsequently perform \emph{policy improvement} \cite[Sec.~4.4]{bertsekas2005dynamic} step to find the policy $\bm{u}^{(k+1)}$ which minimizes the right hand side of Bellman's equation
\begin{equation}
u^{(j+1)}(x) = \arg\min_{u\in \mathcal{U}(x)} \left[g(x, u) + \sum_{y \in \mathcal{S}}p_{x\rightarrow y | u}h^{(j)}(y)\right]. \label{eq:ukplus1}
\end{equation}

If $\bm{u^{(j+1)}} = \bm{u^{(j)}}$, the algorithm terminates, and the optimal policy is obtained $\bm{u^*} = \bm{u^{(j)}}$. Otherwise, repeat the procedure by replacing $\bm{u^{(j)}}$ with $\bm{u^{(j+1)}}$. It is proved that the policy iteration algorithm terminates in finite number of iterations \cite[Prop.~4.4.1]{bertsekas2005dynamic}. To sum up, the policy iteration algorithm is summarized in Algorithm \ref{alg:piter}.

\begin{algorithm}[th]
\caption{Policy Iteration Algorithm} \label{alg:piter}
\begin{algorithmic}[1]

\STATE Set $u^{(1)}(x) = 0$ for all $x \in \mathcal{S}$.

\STATE Set $j = 0$.

\STATE \textbf{Do}

\STATE {~} Set $j = j+1$.

\STATE {~} Calculate $\lambda^{(j)}$ and $\bm{h}^{(j)}$ by solving (\ref{eq:linear}) and (\ref{eq:hconstr}).

\STATE {~} Calculate $\bm{u}^{(j+1)}$ according to (\ref{eq:ukplus1}).

\STATE \textbf{While}($\bm{u}^{(j+1)} \neq \bm{u}^{(j)}$)

\end{algorithmic}
\end{algorithm}

\section{Numerical Results} \label{simulation}

We run some simulations to study the structure of the optimal policy as well as evaluate its performance. We set the cell radius $R=50$m, the required content delivery spectrum efficiency $r_0/W = 1$bps/Hz, the pathloss parameters $\beta = 10$dB and $\alpha = 2$, $T_p = 1$s, $N = 20$, and the Zipf parameter $v = 0.5$. The battery capacity is discretized so that $E_{\mathrm{max}} = 15$, and we set $M=4$, $P_t(R) = 1$Watt, $E_{\mathrm{unit}}$ and $\sigma^2 + I$ are set so that $l_M = M$ and (\ref{eq:rate}) holds for $r=r_0, d = d_M, P_t = P_t(R)$, and then $d_1, \ldots, d_{M-1}$ are selected so that $l_i = i, i = 0, 1, \ldots, M-1$. Assume the energy arrival process follows a Poisson distribution with average arrival rate $\bar{a}$ units of energy.

Fig.~\ref{fig:Optpolicy} shows the optimal policy structure with parameters $p_c = 0.3, p_u = 0.7, \bar{a} = 0.8$. Based on the results, we can have the following observations. Firstly, given user request state and pushed content state, the optimal policy w.r.t.~battery energy state shows a threshold-based structure, i.e., the BS will keep sleep until the battery energy exceeds some value, and then it will not sleep for any battery energy state larger than the value. It is because when the amount of battery energy is large, the BS tends to greedily use it in case of battery overflow. Secondly, for the users close to the BS (user request state 1), unicast is always preferred. As these users experience very good channel quality, unicast consumes very little energy, and hence is more beneficial than transferring the request to the macro BS. Thirdly, the more contents are pushed, the less tendency the system decides to push. For $C_k = 0$, i.e., no contents are pushed, the BS will push the popular contents for most states except that the users are close to the BS. However, when the number of pushed contents approaches its maximum (e.g.~$C_k = 19$), the BS will push only when the system is idle ($Q_k = 0$) and the energy battery is almost full ($E_k \ge 12$).

\begin{figure}
\centering
\includegraphics[width=3.6in]{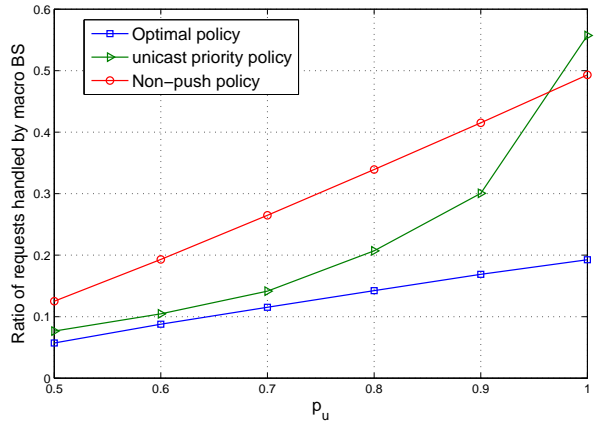}
\caption{The ratio of requests handled by the macro BS with/without proactive push. $p_c = 0.3, \bar{a} = 0.8$.} \label{fig:rvspu}
\end{figure}

Then we evaluate the performance gain obtained by push mechanism, which is illustrated in  Fig.~\ref{fig:rvspu}. Here, the unicast priority policy \cite{Zhou2014greendelivery} is a simple greedy policy in which the BS always satisfy the unicast request in the first place. The push action is taken only when there is no user request. While the non-push policy only takes actions including sleep and unicast, and it is also optimized using MDP approach, which follows the similar procedure of Sec.~\ref{sec:problem} by removing the push action. It is shown that compared with non-push optimal policy, the optimal push mechanism reduces the ratio of requests handled by the macro BS by more than 50\% and the gain increases as the traffic load increases (For the full buffer case where $p_u=1$, the ratio is reduced by 60\%). On the other hand, the unicast priority policy performs close to the optimal push policy at low traffic load regime, but performs even worse than the non-push policy when the traffic load is high. For the low traffic load case, it performs well since there is sufficient idle period for the system to push. While for the high traffic load case, very few contents can be pushed and the unicast priority policy converges to the non-push policy. For the full buffer case, it reduces to greedy non-push policy, and hence performs worse than the optimal non-push policy.

\section{Conclusion} \label{concl}
In this paper, proactive push in EH based SBS is optimized with MDP tools by properly discretizing the system energy and user request states. With policy iteration policy, the optimal policy is found and is shown by numerical results that it reveals a threshold based structure, i.e., the BS sleeps until the battery energy exceeds some threshold. Then it keeps its unicast/push action for the rest battery energy states. In addition, compared with non-push policy, the push based optimal policy reduces the ratio of requests handled by macro BS by more than 50\%. It is shown that the push mechanism has great potential for performance enhancement. As this paper mainly focuses on problem formulation and algorithm design for optimal policy, future work includes the analysis of the structure of the optimal policy. Also, integrating the non-ideal content fetch and caching in SBS is also a potential research direction.


\section*{Acknowledgment}

This work is sponsored in part by the National Basic Research Program of China (No.~2012CB316001), and the Nature Science Foundation of China (No.~61201191 and 61401250), the Creative Research Groups of NSFC (No.~61321061), the Sino-Finnish Joint Research Program of NSFC (No.~61461136004), and Hitachi R\&D Headquarter.

\bibliographystyle{IEEEtran}
\bibliography{ref}

\begin{thebibliography}{10}
\providecommand{\url}[1]{#1}
\csname url@samestyle\endcsname
\providecommand{\newblock}{\relax}
\providecommand{\bibinfo}[2]{#2}
\providecommand{\BIBentrySTDinterwordspacing}{\spaceskip=0pt\relax}
\providecommand{\BIBentryALTinterwordstretchfactor}{4}
\providecommand{\BIBentryALTinterwordspacing}{\spaceskip=\fontdimen2\font plus
\BIBentryALTinterwordstretchfactor\fontdimen3\font minus
  \fontdimen4\font\relax}
\providecommand{\BIBforeignlanguage}[2]{{%
\expandafter\ifx\csname l@#1\endcsname\relax
\typeout{** WARNING: IEEEtran.bst: No hyphenation pattern has been}%
\typeout{** loaded for the language `#1'. Using the pattern for}%
\typeout{** the default language instead.}%
\else
\language=\csname l@#1\endcsname
\fi
#2}}
\providecommand{\BIBdecl}{\relax}
\BIBdecl

\bibitem{ozel2011transmission}
O.~Ozel, K.~Tutuncuoglu, J.~Yang, S.~Ulukus, and A.~Yener, ``Transmission with
  energy harvesting nodes in fading wireless channels: Optimal policies,''
  \emph{IEEE Journal on Selected Areas in Communications}, vol.~29, no.~8, pp.
  1732--1743, 2011.

\bibitem{gunduz2014designing}
D.~Gunduz, K.~Stamatiou, N.~Michelusi, and M.~Zorzi, ``Designing intelligent
  energy harvesting communication systems,'' \emph{IEEE Communications
  Magazine}, vol.~52, no.~1, pp. 210--216, Jan. 2014.

\bibitem{niu2008new}
Z.~Niu, L.~Long, J.~Song, and C.~Pan, ``A new paradigm for mobile multimedia
  broadcasting based on integrated communication and broadcast networks,''
  \emph{IEEE Communications Magazine}, vol.~46, no.~7, pp. 126--132, Jul. 2008.

\bibitem{liu2013utility}
J.~Liu, W.~Chen, Y.~J. Zhang, and Z.~Cao, ``A utility maximization framework
  for fair and efficient multicasting in multicarrier wireless cellular
  networks,'' \emph{IEEE/ACM Transactions on Networking}, vol.~21, no.~1, pp.
  110--120, Feb. 2013.

\bibitem{podnar2002distributed}
I.~Podnar, M.~Hauswirth, and M.~Jazayeri, ``Mobile push: delivering content to
  mobile users,'' in \emph{Proceedings. 22nd International Conference on
  Distributed Computing Systems Workshops}, 2002, pp. 563--568.

\bibitem{zhou2013wireless}
X.~Zhou, R.~Zhang, and C.~K. Ho, ``Wireless information and power transfer:
  Architecture design and rate-energy tradeoff,'' \emph{IEEE Transactions on
  Communications}, vol.~61, no.~11, pp. 4754--4767, November 2013.

\bibitem{golrezaei2013femtocaching}
N.~Golrezaei, A.~F. Molisch, A.~G. Dimakis, and G.~Caire, ``Femtocaching and
  device-to-device collaboration: A new architecture for wireless video
  distribution,'' \emph{IEEE Communications Magazine}, vol.~51, no.~4, pp.
  142--149, Apr. 2013.

\bibitem{Shanmugam2013femtocaching}
K.~Shanmugam, N.~Golrezaei, A.~F. Molisch, A.~G. Dimakis, and G.~Caire,
  ``Femtocaching: wireless content delivery through distributed caching
  helpers,'' \emph{IEEE Transactions on Information Theory}, vol.~59, no.~12,
  pp. 8402--8413, Dec. 2013.

\bibitem{wang2011on}
X.~Wang, Y.~Bao, X.~Liu, and Z.~Niu, ``On the design of relay caching in
  cellular networks for energy efficiency,'' in \emph{IEEE Conference on
  Computer Communications Workshops (INFOCOM WKSHPS)}, Apr. 2011, pp. 259--264.

\bibitem{bastug2014living}
E.~Bastug, M.~Bennis, and M.~Debbah, ``Living on the edge: The role of
  proactive caching in 5g wireless networks,'' \emph{IEEE Communications
  Magazine}, vol.~52, no.~8, pp. 82--89, Aug 2014.

\bibitem{Hiwifi}
\BIBentryALTinterwordspacing
 [Online]. Available: \url{http://www.hiwifi.com/j2}
\BIBentrySTDinterwordspacing

\bibitem{wang2014push}
K.~Wang, Z.~Chen, and H.~Liu, ``Push-based wireless converged networks for
  massive multimedia content delivery,'' \emph{IEEE Transactions on Wireless
  Communications}, vol.~13, no.~5, pp. 2894--2905, May 2014.

\bibitem{sharma2013greencache}
N.~Sharma, D.~Krishnappa, D.~Irwin, M.~Zink, and P.~Shenoy, ``Greencache:
  augmenting off-the-grid cellular towers with multimedia caches,'' in
  \emph{Proc. ACM MMsys¡¯13}, Feb. 2013.

\bibitem{Zhou2014greendelivery}
S.~Zhou, J.~Gong, Z.~Zhou, W.~Chen, and Z.~Niu, ``Greendelivery: Proactive
  content caching and push with energy harvesting based small cells,''
  \emph{submitted to IEEE Communications Magazine}.

\bibitem{bertsekas2005dynamic}
D.~P. Bertsekas, \emph{Dynamic programming and optimal control, Volume II,
  \emph{3rd edition}}.\hskip 1em plus 0.5em minus 0.4em\relax Athena Scientific
  Belmont, MA, 2005.

\bibitem{cha2007tube}
M.~Cha, H.~Kwak, P.~Rodriguez, Y.-Y. Ahn, and S.~Moon, ``I tube, you tube,
  everybody tubes: analyzing the world's largest user generated content video
  system,'' in \emph{Proceedings of the 7th ACM SIGCOMM Conference on Internet
  measurement}.\hskip 1em plus 0.5em minus 0.4em\relax ACM, 2007, pp. 1--14.

\end{thebibliography}

\end{document}